\documentclass[12pt]{iopart}

\usepackage{graphicx}% Include figure files
\usepackage{dcolumn}% Align table columns on decimal point
\usepackage{bm}% bold math

\begin{document}

\title[Optical, transport and magnetic properties of new compound CeCd$_{3}$P$_{3}$]{Optical, transport and magnetic properties of new compound CeCd$_{3}$P$_{3}$}

\author{Shohei Higuchi$^1$, Yuki Noshima$^1$, Naoki Shirakawa$^2$, Masami Tsubota$^3$, and Jiro Kitagawa$^1$}

\address{$^1$ Department of Electrical Engineering, Faculty of Engineering, Fukuoka Institute of Technology, 3-30-1 Wajiro-higashi, Higashi-ku, Fukuoka 811-0295, Japan}
\address{$^2$ National Institute of Advanced Industrial Science and Technology, Tsukuba, Ibaraki 305-8658, Japan}
\address{$^3$ Physonit Inc., 6-10 Minami-Horikawa, Kaita Aki, Hiroshima 736-0044, Japan}
\ead{j-kitagawa@fit.ac.jp}

\begin{abstract}
We have found that CeCd$_{3}$P$_{3}$ crystallizes into a hexagonal ScAl$_{3}$C$_{3}$-type structure. The optical, transport and magnetic properties of CeCd$_{3}$P$_{3}$ were investigated by measuring the diffuse reflectance, electrical resistivity and magnetization. CeCd$_{3}$P$_{3}$ is a semiconductor with the fundamental band gap of approximately 0.75 eV. The 4$f$ electrons of Ce$^{3+}$ ions are well localized but do not show long range order down to 0.48 K, presumably due to the geometrical frustration of Ce atoms. The magnetic ordering temperature is possibly lower than that of isostructural CeZn$_{3}$P$_{3}$ (0.75 K). Because several $f$-electron compounds with the ScAl$_{3}$C$_{3}$-type structure are quantum spin systems, CeCd$_{3}$P$_{3}$ may be a candidate of quantum spin liquid. On the other hand, the relatively large band gap compared to approximately 0.4 eV in CeZn$_{3}$P$_{3}$, would not be intimate with the observation of photoinduced Kondo effect, providing a potentially new range of applications of devices based on the Kondo effect.
\end{abstract}

\vspace{2pc}

%\submitto{\NJP}
\maketitle

\clearpage

\section{Introduction}
Frustrated spin systems have been attracting much attention for long term.
Recent research trend is study of frustrated magnets, showing quantum critical behaviors such as spin liquids and non-Fermi liquids\cite{Balents:Nature2010,Trovarelli:PRL2000}.
In Ce compounds, a crystalline-electric-field (CEF) ground state of Ce trivalent ion possesses a rather large quantum-spin-fluctuation, and pyrochlore Ce$_{2}$Sn$_{2}$O$_{7}$ and ScAl$_{3}$C$_{3}$-type Ce-compounds are quantum spin systems\cite{Sibille:PRL2015,Yamada:JPCS2010,Ochiai:JPCMconf2010,Hara:PRB2012}.
Magnetic properties of Ce$_{2}$Sn$_{2}$O$_{7}$ are governed by a Kramers doublet, which can be regarded as effective $S$=1/2 spin.
Ce$_{2}$Sn$_{2}$O$_{7}$ does not show long range magnetic ordering down to 0.02 K due to the geometrical frustration of Ce atoms associated with the pyrochlore structure, which suggests a quantum-spin-liquid behavior\cite{Sibille:PRL2015}.

\begin{figure}
\begin{center}
\includegraphics[width=12cm]{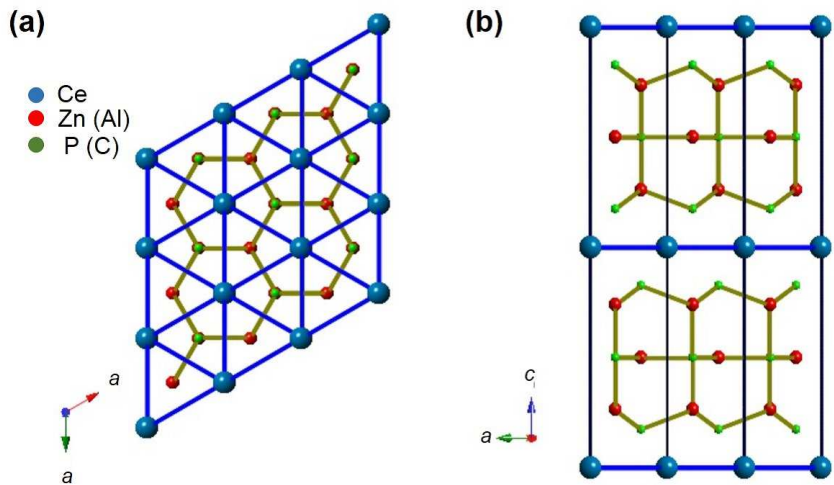}
\end{center}
\caption{Crystal structure of ScAl$_{3}$C$_{3}$-type Ce compounds. (a) 2D plane formed by Ce (blue sphere) triangular lattice perpendicular to $c$-axis. (b) Stacking Ce-layers along $c$-axis. The red and green spheres represent Zn (Al) and P (C) atoms, respectively.}
\label{f1}
\end{figure}

Semiconducting\cite{Yamada:JPCS2010,Kitagawa:JPSJ2013,Kitagawa:PRB2016} CeZn$_{3}$P$_{3}$ and metallic\cite{Ochiai:JPCMconf2010} CeAl$_{3}$C$_{3}$ are the typical member of hexagonal ScAl$_{3}$C$_{3}$-type\cite{Nientiedt:JSSC1999} Ce-compounds, the crystal structure of which is shown in Fig.\ 1.
Ce atoms form the triangular lattice of the two-dimensional layer, sandwiching Zn(Al) and P(C) atoms. 
Curie-Weiss law behaviors of both compounds in the magnetic susceptibility indicate localized 4$f$ spins of the Ce$^{3+}$ ions\cite{Yamada:JPCS2010,Ochiai:JPCMconf2010}. 
Reflecting the frustrated Ce atom geometry, CeZn$_{3}$P$_{3}$ has a low magnetic ordering temperature\cite{Yamada:JPCS2010} of 0.75 K. 
A dimerization of 4$f$ spins is observed at approximately 4 K in the metallic CeAl$_{3}$C$_{3}$, which is classified as a quantum spin system\cite{Ochiai:JPCMconf2010}.

In 4$f$ metallic systems, the magnetic interaction competes with the Kondo one.
The latter interaction originates from the screening of 4$f$ magnetic moment by a conduction spin one.
In accordance with the so-called Doniach phase diagram\cite{Hewson:book1993,Brandt:AP1984} for a compound with weakened magnetic interaction between 4$f$ spins, the Kondo interaction dominates the system.
Based on the idea, we have recently found the photoinduced Kondo effect\cite{Kitagawa:PRB2016} in CeZn$_{3}$P$_{3}$.
The Kondo effect emerges under visible-light illumination of the material through the photocarrier doping.
The photoinduced Kondo effect provides a potentially new range of operation for not only quantum information/computation devices but also operation of magneto-optic devices, thereby expanding the range of applications of devices based on the Kondo effect.  

Nientiedt et al. have reported PrCd$_{3}$P$_{3}$ crystallizing into the ScAl$_{3}$C$_{3}$-type structure\cite{Nientiedt:JSSC1999}.
We tried replacing the Pr atom with the Ce atom, expecting a new compound showing a quantum spin liquid and/or a photoinduced Kondo effect.
If CeCd$_{3}$P$_{3}$ exists, the magnetic ordering temperature would be lower than that of CeZn$_{3}$P$_{3}$, because the ionic radius of Cd ion is larger than that of Zn ion.
The lowered ordering temperature is favorable for the realization of quantum spin liquid and/or photoinduced Kondo effect.
We have found that CeCd$_{3}$P$_{3}$ is a new member of the ScAl$_{3}$C$_{3}$-type structure.
To easily compare experimental optical and transport properties with a band calculation result, we have synthesized the non-magnetic counterpart LaCd$_{3}$P$_{3}$, which is also a new member of the ScAl$_{3}$C$_{3}$-type structure.
In this paper, we report the synthesis and characterization of polycrystalline samples.
The optical, transport and magnetic properties were investigated by measuring the diffuse reflectance, electrical resistivity and magnetization.

\section{Materials and Methods}

\begin{figure}
\begin{center}
\includegraphics[width=11cm]{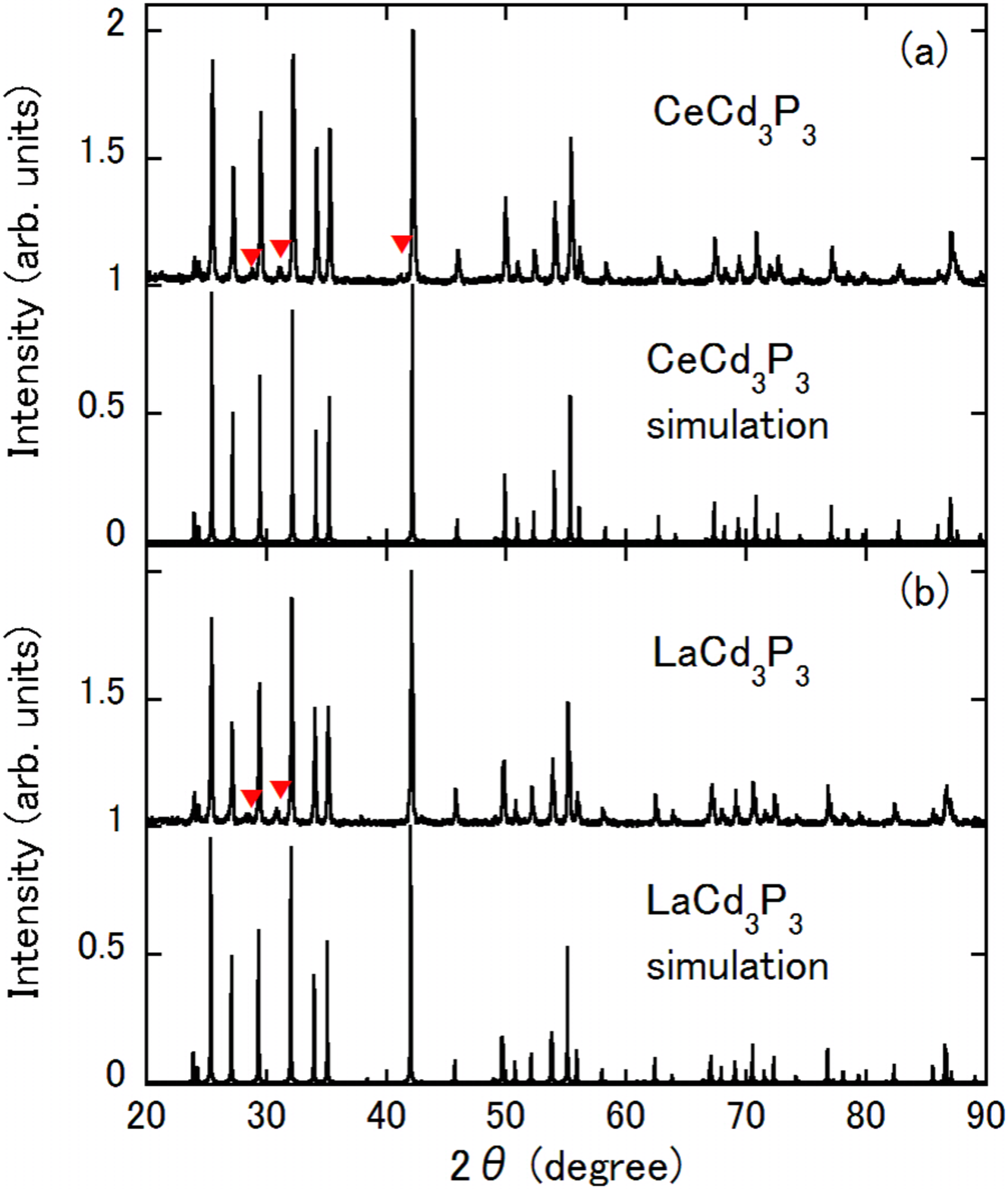}
\end{center}
\caption{(a) XRD pattern of CeCd$_{3}$P$_{3}$. The simulated pattern is also shown. The origin of each pattern is shifted by an integer value for clarity. The triangles show the parasitic oxide phase. (b) XRD pattern of LaCd$_{3}$P$_{3}$. The simulated pattern is also shown. The origin of each pattern is shifted by an integer value for clarity. The triangles show the parasitic oxide phase.}
\label{f2}
\end{figure}

\begin{figure}
\begin{center}
\includegraphics[width=9cm]{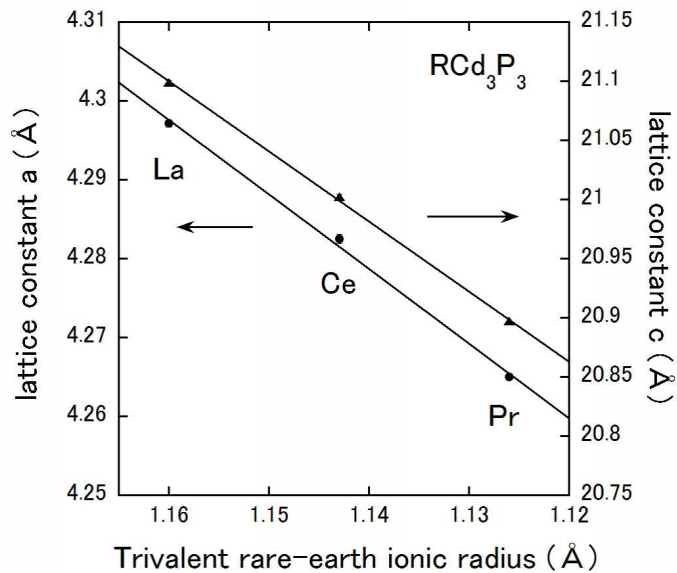}
\end{center}
\caption{The lattice parameters of RCd$_{3}$P$_{3}$ (R=La, Ce and Pr) are plotted against the trivalent ionic radius of rare-earth ion proposed by Shannon\cite{Shannon:AC1976}. The lanthanide contraction holds fairly well. The lattice parameters of PrCd$_{3}$P$_{3}$ are referred from \cite{Nientiedt:JSSC1999}.}
\label{f3}
\end{figure}

Polycrystalline CeCd$_{3}$P$_{3}$ and LaCd$_{3}$P$_{3}$ samples were synthesized by a solid-state reaction technique using Ce (or La) pieces (99.9\%), Cd pieces (or powder) (99.9999\% or (99.9\%)) and red P powder (99.999\%).
We used CeCd$_{2}$ or LaCd$_{2}$, prepared by direct reaction of the constituent elements in an evacuated quartz tube that was heated at 900 $^{\circ}$C for 1 h, as the solid-state reaction precursor. 
Crushed CeCd$_{2}$ (LaCd$_{2}$), Cd powder and P with the molar ratio CeCd$_{2}$ (LaCd$_{2}$):Cd:P = 1:1:3 were homogeneously mixed together in a glove box. 
The pelletized sample was finally reacted in an evacuated quartz tube at 800 $^{\circ}$C for 2 days.
The samples were evaluated using a powder X-ray diffractometer with Cu-K$\alpha$ radiation. 

We checked the optical band gaps of CeCd$_{3}$P$_{3}$ and LaCd$_{3}$P$_{3}$ via their diffuse reflectance $R_{d}$ spectra, recorded using a UV-Vis spectrometer (Shimadzu, UV-3600). 
The temperature dependence of electrical resistivity $\rho$(T) between 20 and 300 K was measured by the conventional DC four-probe method using a closed-cycle He gas cryostat.
The temperature dependence of DC magnetization $\chi$(T) from 0.48 K to 300 K under a magnetic field of 100 Oe was measured using a Quantum Design MPMS equipped with iHelium3\cite{Shirakawa:Poly2005}.

The band calculation was performed by the linearized augmented plane wave (LAPW) method using the WIEN2K package\cite{Blaha:book2001}. 
The exchange-correlation potential was described using generalized gradient approximations with the Perdew-Burke-Ernzerhof (PBE) functional. 
For the calculation, we used 1000 $k$-points ($k$:wave number) and a plane wave cut-off parameter, $RK_{max}$, of 7.

\section{Results and discussion}
The X-ray diffraction (XRD) patterns of CeCd$_{3}$P$_{3}$ and the La counterpart are shown in Figs.\ 2(a) and 2(b), respectively.
They were well indexed by the hexagonal space group $P6_{3}/mmc$ (194) except for the peaks denoted by triangles.
The lattice parameters were refined by the least square method as $a$=4.28251(62) \AA \hspace{1.5mm} and $c$=21.00239(230) \AA \hspace{1.5mm} for CeCd$_{3}$P$_{3}$ ($a$=4.29715(48) \AA \hspace{1.5mm} and $c$=21.09883(180) \AA \hspace{1.5mm} for LaCd$_{3}$P$_{3}$).
Each simulated pattern was calculated by using the lattice parameters and the atomic coordinates of PrCd$_{3}$P$_{3}$ in the literature\cite{Nientiedt:JSSC1999}, and closely matches the experimental pattern.
Each sample contains the small amount of impurity phase (see triangles in Figs.\ 2(a) and 2(b)) identified as the oxide\cite{Orlova:Kri2005} Ce$_{2}$Cd$_{0.5}$(PO$_{4}$)$_{3}$ (or the La counterpart), which cannot be completely removed by trying various synthesis conditions.
In Fig.\ 3, the lattice parameters of RCd$_{3}$P$_{3}$ (R=La,Ce and Pr) are plotted as a function of ionic radius for a trivalent state\cite{Shannon:AC1976}.
The data points generally follow well the lanthanide contraction of the free trivalent R ions (shown as the straight lines in the Figure). 

\begin{figure}
\begin{center}
\includegraphics[width=11cm]{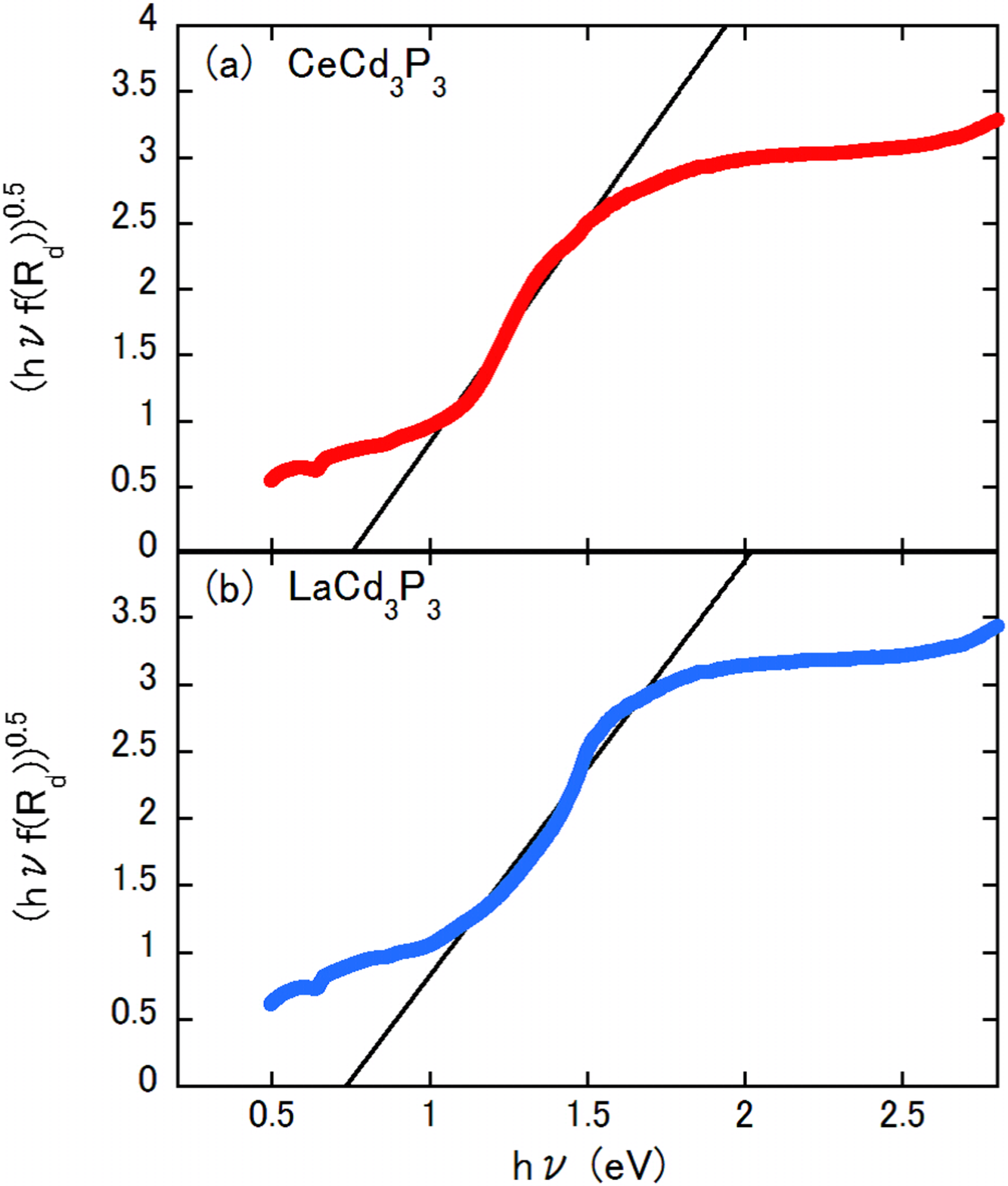}
\end{center}
\caption{Spectra of $(h\nu f(R_{d}))^{1/2}$ of (a) CeCd$_{3}$P$_{3}$ and (b) LaCd$_{3}$P$_{3}$.}
\label{f4}
\end{figure}

The band gap $E_{g}$ can be extracted from the diffuse reflectance $R_{d}$ data by using the so-called Tauc plot\cite{Tauc:PSS1966}.
First of all, the Kubelka-Munk function\cite{Marx:PRB1984,Kozasa:MRE2014} $f(R_{d})$, corresponding to the absorption coefficient, is calculated using the following equation,
\begin{equation}
 f(R_{d})=\frac{(1-R_{d})^2}{2R_{d}}.
\label{equ:KM-func}
\end{equation}
Then $f(R_{d})$ is related to the band gap $E_{g}$ through 
\begin{equation}
(h\nu f(R_{d}))^{1/m}=A(h\nu-E_{g}), 
\label{equ:Tauc relation}
\end{equation}
 where $h$, $\nu$, and $A$ are the Planck constant, the frequency of the light, and a proportionality factor, respectively\cite{Tauc:PSS1966}.
When $(h\nu f(R_{d}))^{1/m}$ is plotted as a function of $h\nu$ (Tauc plot), $E_{g}$ can be extracted from the horizontal intercept of tangential line along the linear portion of the Tauc plot.
The value of $m$ depends on the optical transition type. 
We used $m$=2, following the report on the indirect-gap isostructural-semiconductor\cite{Tejedor:JCG1995} SmZn$_{3}$P$_{3}$, and also the ab-initio band calculation results for LaCd$_{3}$P$_{3}$.
Figures 4(a) and 4(b) show the spectra of $(h\nu f(R_{d}))^{1/2}$ of CeCd$_{3}$P$_{3}$ and the La counterpart, respectively.
The spectra suggest that both compounds are semiconductor. 
As shown in the Figures, $E_{g}$ of 0.75 (0.73) eV was determined using the tangential line for CeCd$_{3}$P$_{3}$ (LaCd$_{3}$P$_{3}$). 
For each compound, the spectral weight at lower energies deviating upward from the solid line indicates the existence of impurity states in the energy gap.

\begin{figure}
\begin{center}
\includegraphics[width=9cm]{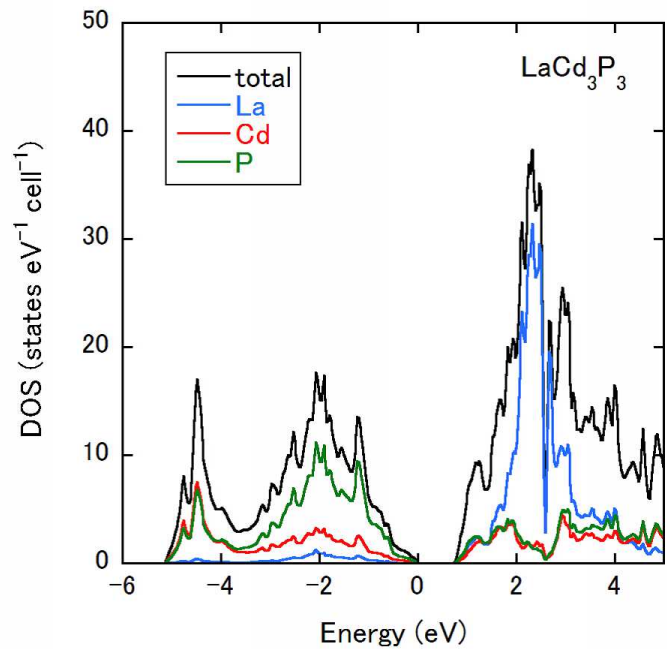}
\end{center}
\caption{Density of states per unit cell of LaCd$_{3}$P$_{3}$. The black curve represents the total density of states. The blue, red and green curves represent the projected density of states of the individual La, Cd, and P atoms, respectively.}
\label{f5}
\end{figure}

Fig.\ 5 shows the total density of states (DOS) profile of LaCd$_{3}$P$_{3}$ with the Fermi energy ($E_{F}$) located at 0 eV. 
The projected DOS profiles for the La, Cd and P atoms are also given. 
The DOS was calculated by employing the lattice parameters obtained from the XRD data.
The atomic coordinates were those of PrCd$_{3}$P$_{3}$\cite{Nientiedt:JSSC1999}.We note here that the atomic coordinates of LaCd$_{3}$P$_{3}$ obtained from a preliminary Rietveld analysis are almost identical to those of PrCd$_{3}$P$_{3}$.
$E_{g}$ was estimated to be 0.75 eV, which agrees well with the experimentally determined value from the $R_{d}$ spectrum. 
The valence and conduction bands are primarily composed of orbitals with covalent bond characteristics between the Cd and P atoms and the La 5$d$ orbital, respectively. 

\begin{figure}
\begin{center}
\includegraphics[width=9cm]{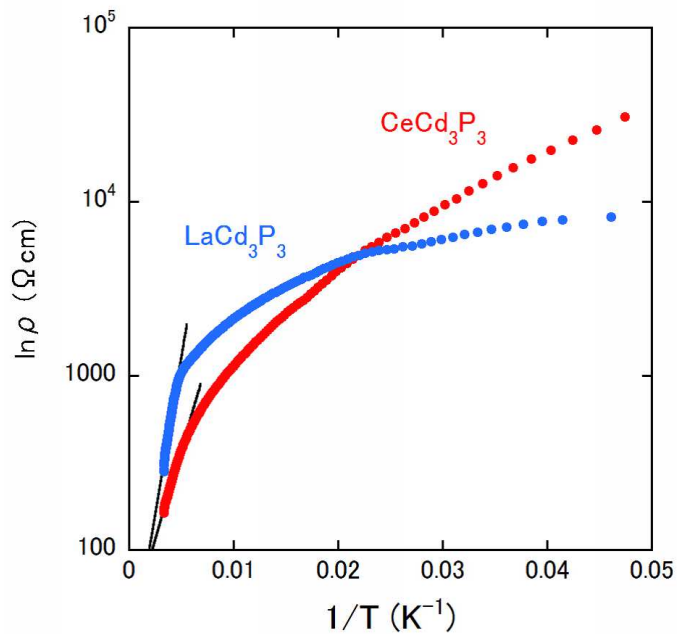}
\end{center}
\caption{Arrhenius plots of $\rho$(T) of CeCd$_{3}$P$_{3}$ and LaCd$_{3}$P$_{3}$.}
\label{f6}
\end{figure}

The temperature dependence of electrical resistivity $\rho$(T) of CeCd$_{3}$P$_{3}$ and the La counterpart show semiconducting behaviors, and the Arrhenius plots of $\rho$(T) are presented in Fig.\ 6.
The activation energy $E_{ac}$ was estimated on the high-temperature side using the relation $\ln\rho\propto E_{ac}/2T$ (see the lines in Fig.\ 6). 
$E_{ac}$ was determined to be about 81 meV (140 meV) for CeCd$_{3}$P$_{3}$ (LaCd$_{3}$P$_{3}$), which is smaller than the fundamental band gap $E_{g}$.
The difference is presumably due to the existence of impurity states in the fundamental band gap.
The semiconducting ScAl$_{3}$C$_{3}$-type Ce-compounds represented by CeZn$_{3}$P$_{3}$ are superior platforms for photoinduced Kondo effect\cite{Kitagawa:PRB2016}, which would require a band gap closing under illumination.
Thus, lower $\rho$ with a smaller $E_{g}$ would be advantageous for the observation of photoinduced Kondo effect.
However $E_{g}$ ($\sim$ 0.75 eV) of CeCd$_{3}$P$_{3}$ is larger than that ($\sim$ 0.4 eV\cite{Kitagawa:JPSJ2013,Kitagawa:PRB2016}) of CeZn$_{3}$P$_{3}$ and consequently $\rho$(T) of CeCd$_{3}$P$_{3}$ tends to be higher.
Therefore we speculate that CeCd$_{3}$P$_{3}$ is not a good candidate showing the photoinduced Kondo effect.

\begin{figure}
\begin{center}
\includegraphics[width=11cm]{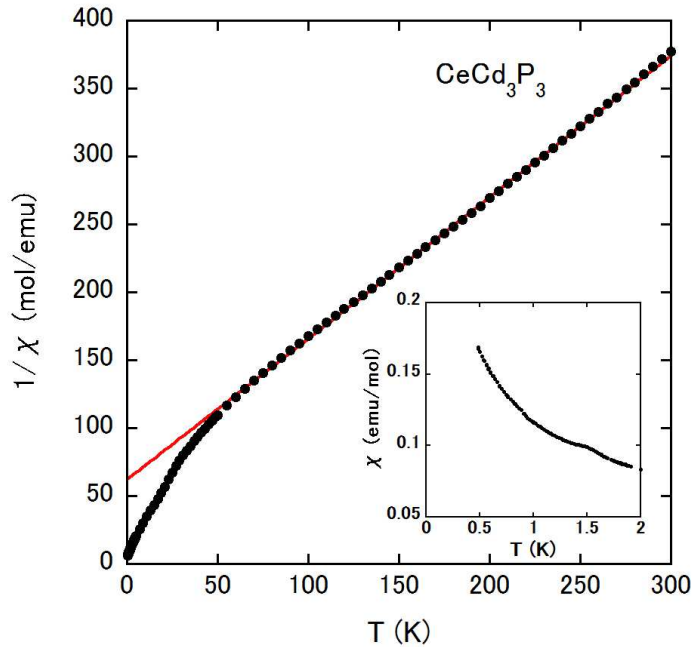}
\end{center}
\caption{Temperature dependence of inverse magnetic susceptibility of CeCd$_{3}$P$_{3}$. The inset shows the low-temperature $\chi$ of the sample.}
\label{f7}
\end{figure}

Figure 7 shows the temperature dependence of the reciprocal $\chi$ of CeCd$_{3}$P$_{3}$.
It follows the Curie-Weiss law above 50 K (see the line in Fig.\ 7).
The effective moment $\mu_{eff}$ and Weiss temperature $\Theta$ are 2.77 $\mu_{B}$/Ce and -60 K, respectively.
The value of $\mu_{eff}$ is near to 2.54 $\mu_{B}$/Ce, expected for a free trivalent Ce ion.
The fact is consistent with the lanthanide contraction as shown in Fig.\ 3.
The inset of Fig.\ 7 is $\chi$(T) below 2 K, which exhibits no magnetic ordering down to 0.48 K.
The faint kink at approximately 1.5 K may be due to the existence of magnetic impurity phase.
We need further purification of the sample.
 
The replacement of Zn atom with Cd one leads to the expansion of lattice parameters ($a$=4.28251(62) \AA, $c$=21.00239(230) \AA \hspace{1.5mm} for CeCd$_{3}$P$_{3}$ and $a$=4.051 \AA, $c$=20.019 \AA \hspace{1.5mm} for CeZn$_{3}$P$_{3}$\cite{Kitagawa:PRB2016}).
Therefore, as expected in the introduction, magnetic ordering temperature of 0.75 K in CeZn$_{3}$P$_{3}$ is possibly reduced below 0.48 K in CeCd$_{3}$P$_{3}$.
Taking into account that several $f$-electron compounds with the ScAl$_{3}$C$_{3}$-type structure are quantum spin systems\cite{Ochiai:JPCMconf2010,Hara:PRB2012}, we may propose that CeCd$_{3}$P$_{3}$ is a candidate of quantum spin liquid.
Further experiments at lower temperature are necessary to confirm the proposal.
The rather large negative $\Theta$-value may be the signature of quantum spin frustration\cite{Balents:Nature2010}.
However in rare-earth compounds, the CEF effect plays a major role for determining $\Theta$-value.
No observation of broad maximum in $\chi$(T) means that a dimerization of 4$f$ spins like in metallic CeAl$_{3}$C$_{3}$ does not occur, which might be common characteristic in semiconducting ScAl$_{3}$C$_{3}$-type Ce-compounds.

\begin{figure}
\begin{center}
\includegraphics[width=9cm]{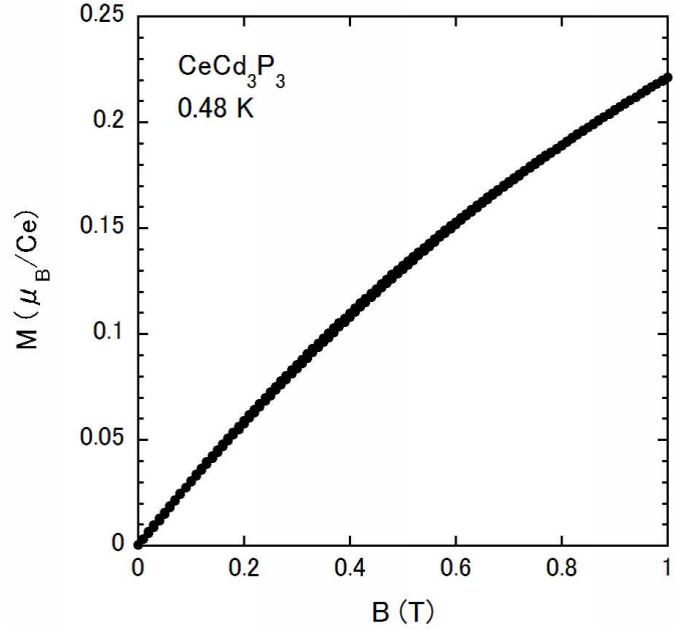}
\end{center}
\caption{Magnetization curve of CeCd$_{3}$P$_{3}$ measured at 0.48 K.}
\label{f8}
\end{figure}

As shown in Fig.\ 8, the magnetization $M$ at 0.48 K shows a paramagnetic behavior up to 1 T, which supports the extrinsic nature of kink of $\chi$(T) at approximately 1.5 K.
The point symmetry of Ce site is trigonal $D_{3d}$.
The CEF Hamiltonian\cite{Alekseev:JETP1994,Kohgi:PhysicaB1993} can be expressed as follows:
\begin{equation}
H_{CEF}=B_{2}^{0}\hat{O}_{2}^{0}+B_{4}^{0}\hat{O}_{4}^{0}+B_{4}^{3}\hat{O}_{4}^{3},
\label{equ:trigonal CEF}
\end{equation}
where ${B}_{n}^{m}$ and $\hat{O}_{n}^{m}$ are the CEF parameters and Stevens' operator equivalents, respectively.
Under the trigonal CEF, the 6-fold degeneracy of the $J$ = 5/2 multiplet of Ce$^{3+}$ ions splits into three doublets:
\begin{equation}
\alpha\left|\pm\frac{1}{2}\right>+\beta\left|\mp\frac{5}{2}\right>, \left|\pm\frac{3}{2}\right>, \beta\left|\pm\frac{1}{2}\right>-\alpha\left|\mp\frac{5}{2}\right>.
\label{equ:CEF wf}
\end{equation}
The $M$ value calculated for the isolated $\left|\pm\frac{3}{2}\right>$ ground state far exceeds 0.225 $\mu_{B}$/Ce at 1 T.
The mixed states between $\left|\pm\frac{1}{2}\right>$ and $\left|\mp\frac{5}{2}\right>$ can reduce the $M$-value.
Therefore $\alpha\left|\pm\frac{1}{2}\right>+\beta\left|\mp\frac{5}{2}\right>$ or $\beta\left|\pm\frac{1}{2}\right>-\alpha\left|\mp\frac{5}{2}\right>$ state would be the CEF ground state.

Here we note that the point symmetry $D_{3d}$ of Ce site in CeCd$_{3}$P$_{3}$ is the same as that in Ce$_{2}$Sn$_{2}$O$_{7}$ which is reported to be the candidate of quantum spin liquid.
The inverse of $\chi$(T) of Ce$_{2}$Sn$_{2}$O$_{7}$ follows a Curie-Weiss law explained by free trivalent Ce ions with large negative $\Theta$ of -220 K\cite{Sibille:PRL2015}.
The magnetization curve\cite{Sibille:PRL2015} of Ce$_{2}$Sn$_{2}$O$_{7}$ shows a paramagnetic behavior and $M$ ($B$=1 T) at 0.5 K reaches a value similar to that of CeCd$_{3}$P$_{3}$.
The CEF ground state of Ce$_{2}$Sn$_{2}$O$_{7}$ is also a linear combination of $\left|\pm\frac{1}{2}\right>$ and $\left|\mp\frac{5}{2}\right>$ states.
The similar CEF ground states under the $D_{3d}$ point symmetry would lead to the common magnetic properties.
Therefore it should be noted that the point symmetry $D_{3d}$ might play a crucial role to realize a quantum spin liquid.

\section{Conclusions}
We have found the semiconductor CeCd$_{3}$P$_{3}$ crystallizing into the hexagonal ScAl$_{3}$C$_{3}$-type structure, possessing geometrically frustrated Ce-triangular lattice.
The physical properties were investigated by measuring $R_{d}$, $\rho$ and $\chi$.
The fundamental band gap is determined to be approximately 0.75 eV, which is larger than that of CeZn$_{3}$P$_{3}$.
The localized Ce moments do not show long range ordering down to 0.48 K.
The magnetic properties of CeCd$_{3}$P$_{3}$ are reminiscent of those of Ce$_{2}$Sn$_{2}$O$_{7}$ which is the candidate of quantum spin liquid.
The magnetic ordering temperature of CeCd$_{3}$P$_{3}$ is possibly lower than that of CeZn$_{3}$P$_{3}$, suggesting that CeCd$_{3}$P$_{3}$ is a promising new member of quantum spin liquid.
The rather large band gap of CeCd$_{3}$P$_{3}$ would not favor the occurrence of photoinduced Kondo effect.

\ack
J.K. is grateful for the financial support provided by the Asahi Glass Foundation and Comprehensive Research Organization of Fukuoka Institute of Technology.

\end{document}